\documentclass[english,twocolumn,aps,prc,superscriptaddress]{revtex4}
\usepackage[T1]{fontenc}
\usepackage[latin1]{inputenc}
\usepackage{graphicx}
\usepackage{amssymb}

\makeatletter

\providecommand{\tabularnewline}{\\}

\makeatletter

\makeatletter

\makeatletter

\makeatletter

\makeatletter

\makeatletter

\makeatletter

\makeatletter

\makeatletter



\makeatother

\makeatother

\makeatother

\makeatother

\makeatother

\makeatother

\makeatother

\makeatother

\makeatother

\usepackage{babel}
\makeatother
\begin{document}

\title{Spectroscopy of $^{9}$C via resonance scattering of protons on $^{8}$B.}

\author{G.V. Rogachev}

\email{grogache@fsu.edu}

\affiliation{Department of Physics, Florida State University, Tallahassee, FL,
32306}

\author{J.J. Kolata}

\affiliation{Department of Physics, University of Notre Dame, Notre Dame, IN,
46556}

\author{A.S. Volya}

\affiliation{Department of Physics, Florida State University, Tallahassee, FL,
32306}

\author{F.D. Becchetti}

\affiliation{Department of Physics and Astronomy, University of Michigan, Ann
Arbor, MI, 48109}

\author{Y. Chen}

\affiliation{Department of Physics and Astronomy, University of Michigan, Ann
Arbor, MI, 48109}

\author{P.A. DeYoung}

\affiliation{Department of Physics, Hope College, Holland, MI, 49422}

\author{J. Lupton}

\affiliation{Department of Physics and Astronomy, University of Michigan, Ann
Arbor, MI, 48109}

\begin{abstract}
The structure of the neutron-deficient $^{9}$C isotope was studied
via elastic scattering of radioactive $^{8}$B on protons. An excitation
function for resonance elastic scattering was measured in the energy
range from 0.5 to 3.2 MeV in the center-of-momentum system. A new
excited state in $^{9}$C was observed at an excitation energy of
3.6 MeV. An R-matrix analysis indicates spin-parity 5/2$^{-}$ for
the new state. The results of this experiment are compared with Continuum
Shell Model calculations. 
\end{abstract}
\maketitle

\section{introduction}

Light nuclei play a central role in nuclear physics since they are
the simplest cases where nuclear many-body dynamics can be understood
and explored. This special role is reflected in the number of different
theoretical approaches that exist, and overlap, in addressing the
structure of light nuclei. The techniques range from full \textit{ab
initio} methods, such as the Green's Function Monte Carlo method \cite{Pieper2002}
or the Large Basis No-Core Shell Model method \cite{Navratil1998},
in which the properties of light nuclei are computed starting from
bare nucleon-nucleon interactions, to the traditional shell-model
approach with renormalized or phenomenologically-determined interactions,
to cluster models. Light nuclei also provide an important arena for
exploring physics on the interface between structure and reactions.
The dawn of modern structure-reaction unification is marked by several
recent theoretical developments: the Gamow Shell Model (GSM) \cite{Michel2003}
and Continuum Shell Model (CSM) \cite{Volya2005} in particular. Realistic
tests of these models constitute an important step both in theoretical
developments in the description of the properties of exotic nuclei
and in furthering our understanding of the role that exotic resonances
play in nuclear astrophysics.

In the analysis of the experimental study reported below we use the
CSM approach which permits calculation of resonance parameters and
cross sections, thus allowing a direct comparison with experimental
data. A comparison with older techniques such as a traditional R-matrix
analysis with shell-model spectroscopic factors highlights their limitations
and gives a valuable insight into physics on the reaction/structure
borderline encompassed by the advanced CSM technique.

The focus of this work is the structure of an exotic neutron-deficient
dripline isotope of carbon, $^{9}$C, which has a halflife of 126.5
ms and a binding energy of 1.3 MeV. Spectroscopic information about
this nucleus is scarce. The ground state of $^{9}$C was first identified
in the $^{12}$C($^{3}$He,$^{6}$He) reaction by Cerny, \textit{et
al.} \cite{CERNY1964} in 1964. Ten years later, the first excited
state of $^{9}$C was observed at 2.2 MeV using the same reaction
\cite{BENENSON1974}. More recently, $^{12}$C($^{3}$He,$^{6}$He)
was yet again studied in Ref. \cite{GOLOVKOV1991}. The authors of
this work claim to have seen another excited state at about 3.3 MeV.
However, the complicated nature of the reaction mechanism did not
allow for a reliable spin-parity assignment. Instead, indirect arguments
were given favoring spin-parity of 5/2$^{+}$ for this level \cite{GOLOVKOV1991}.

Modern experimental techniques allow for the use of beams of radioactive
nuclei to populate states in exotic isotopes by means of simple reactions
such as one-nucleon transfer or resonance elastic scattering. As a
result, more reliable identification on the properties of the observed
states can now be obtained. In the present work, excited states in
$^{9}$C are populated via resonance elastic scattering of radioactive
$^{8}$B on protons. The main goal of this work was to identify the
excited states in $^{9}$C and compare the properties of these states
with theoretical predictions.

\section{Experiment}

\begin{figure}
\includegraphics[width=1\columnwidth]{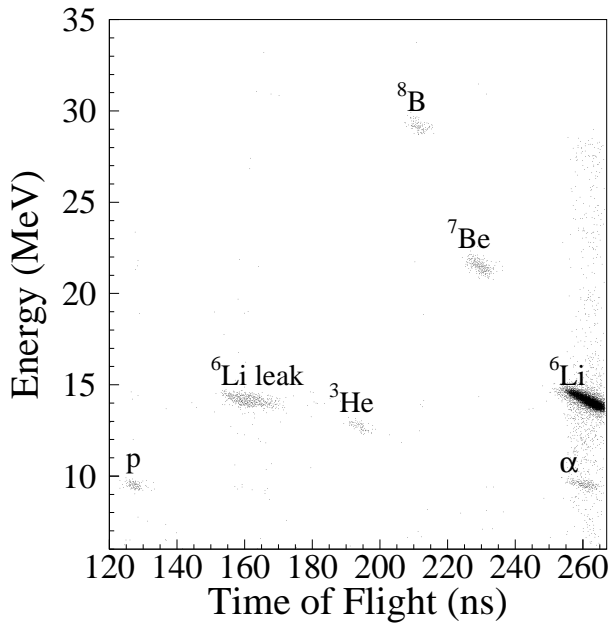}

\caption{\label{fig:tof} The total energy of a beam ion \textit{vs.} its
time of flight measured with the Au target by a Si $\Delta$E-E telescope
placed at 7.7$^{\circ}$. The start signal was taken from the E detector
of the telescope and the stop signal came from the beam buncher.}
\end{figure}

The experiment was performed using the \textit{TwinSol} radioactive
nuclear beam (RNB) facility at the University of Notre Dame \cite{Lee1999}.
A 2.5 cm long gas target containing 1 atm of $^{3}$He was bombarded
by a nanosecond-bunched primary $^{6}$Li beam at an energy of 39.0
MeV and intensity of 200 electrical nA. The two-proton pickup reaction
$^{3}$He($^{6}$Li,$^{8}$B)n was used to produce the $^{8}$B ions.
The entrance and exit windows of the gas cell consisted of 2.0 $\mu$m
Havar foils. The secondary $^{8}$B beam was momentum selected and
transported through the two superconducting solenoids, which focused
it into a 5 mm spot on a 9.0 mg/cm$^{2}$ plastic (CH$_{2}$) target.
The laboratory energy of the $^{8}$B beam at the secondary target
position was 29 MeV, with a resolution of 0.7 MeV full width at half
maximum (FWHM) and an intensity of up to 10$^{4}$ particles per second.
The energy spread was due to a combination of the kinematic shift
in the production reaction plus energy-loss straggling in the gas-cell
windows. Contaminant ions which had the same magnetic rigidity as
29 MeV $^{8}$B also were present in the beam, but they could be identified
using the time-of-flight (TOF) technique. The TOF of the particles
was obtained from the time difference between the occurrence of an
E signal in a detector telescope and the RF timing pulse from the
beam buncher. The time resolution of better than 5 ns (FWHM) was adequate
to cleanly separate $^{8}$B from all other ions (except for some
direct protons as will be discussed further in the text). This is
illustrated in Fig. \ref{fig:tof}, which was obtained with a 1.0
mg/cm$^{2}$ Au target and a Si $\Delta$E-E telescope placed at 7.7$^{\circ}$
with respect to the beam axis. The intensity of the beam during the
experiment was determined from the ratio of the $^{8}$B ions to the
integrated charge of the primary $^{6}$Li beam collected in the \textit{TwinSol}
Faraday cup. This ratio was measured by placing the Si $\Delta$E-E
telescope directly at the target position.

The thick-target inverse-kinematics technique \cite{Artemov1990}
was used to measure the excitation function of $^{8}$B+p resonance
elastic scattering. The plastic target was thick enough (9.0 mg/cm$^{2}$)
to stop the 29 MeV $^{8}$B ions. The recoil protons from back-angle
elastic scattering in this target lose only a small amount of energy
in traversing the foil and emerge from it with sufficient energy to
be detected. Note that the lowest-energy protons, from the scattering
of $^{8}$B ions near the end of their range, encounter the least
amount of material before leaving the target. In this manner, an excitation
function for elastic scattering down to very low energies can be measured
with high efficiency and good energy resolution. The recoil protons
were detected with two telescopes consisting of 19.5 and 19.2 $\mu$m
Si $\Delta$E detectors, backed by 1000 $\mu$m Si E detectors. The
active area of the $\Delta$E detectors was 450 mm$^{2}$, and that
of the E detectors was 600 mm$^{2}$. Each telescope had a circular
collimator with a diameter of 18 mm that subtended a solid angle of
11.6 msr. They were placed on either side of the beam at 7.7$^{\circ}$
with respect to the beam axis. It would have been preferable to place
a telescope at 0$^{\circ}$ to the beam but the light-ion contamination
(Fig. \ref{fig:tof}) produced a count rate in this position that
was unacceptable since these ions penetrated the target and directly
entered the telescope. 

\begin{figure}
\includegraphics[width=1\columnwidth]{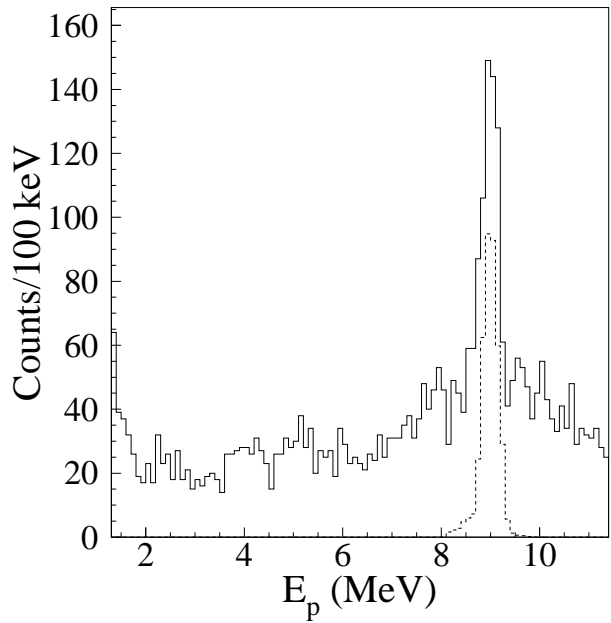}

\caption{\label{fig:raw}The raw spectrum of protons measured by one of the
telescopes and gated on $^{8}$B TOF. The sharp peak at 9 MeV is due
to direct protons scattered by the plastic target. The dashed line
is a spectrum of protons produced by gating on the TOF between $^{8}$B
and $^{7}$Be. }
\end{figure}

It was verified that the recoil proton TOF signal was only slightly
shifted in time relative to $^{8}$B and was stable during the course
of the experiment so that the separation from elastic scattering of
contaminant ions was excellent. Nevertheless, a background associated
with direct protons scattered by the plastic target was still present
in the proton spectrum measured by the telescopes, even after the
TOF gate. The origin of this proton background is the unfortunate
coincidence between the difference of the actual flight times of $^{8}$B
and protons from the primary to the secondary target ($\sim$80 ns)
and the half-period of the buncher (100 ns). Because of this coincidence,
a small fraction of the direct protons from the tail of the proton
time distribution overlaped with the timing of the $^{8}$B ions from
the {}``previous'' bunch. This background was reduced by two orders
of magnitude after a pulse selection was introduced (the efficiency
of pulse selection was 99\%), making the period between beam bunches
200 ns. Still, some direct protons from the tail of the proton time
distribution leaked into the gate, as shown in Fig. \ref{fig:raw}
which represents the raw proton spectrum measured by one of the telescopes
gated by the $^{8}$B TOF. This background was eliminated by subtracting
the spectrum of protons gated on timing in the region between $^{8}$B
and $^{7}$Be (dashed line in Fig. \ref{fig:raw}) from the spectrum
of protons associated with $^{8}$B. %

\begin{figure}
\includegraphics[width=1\columnwidth]{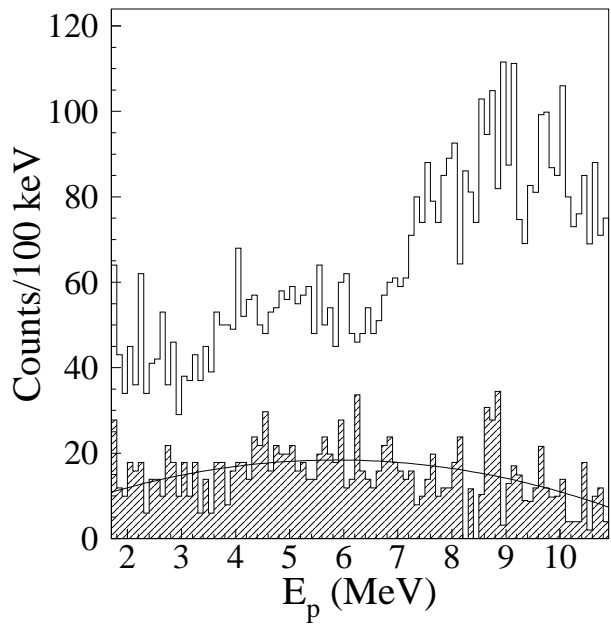}

\caption{\label{fig:total}The laboratory spectrum of protons from both telescopes
measured with the plastic target. The background associated with the
direct protons from the beam has been subtracted. The normalized proton
background from the interaction of $^{8}$B with carbon in the plastic
target is shown as the shaded histogram. The solid line represents
the polynomial fit to the carbon background which was used for subtraction
of this background from the total spectrum.}
\end{figure}

Another source of background are protons produced in the interaction
of the $^{8}$B with carbon in the plastic target. These protons have
exactly the same timing as the recoil protons from elastic scattering
of $^{8}$B on hydrogen. The spectrum of this process, measured using
a 15.8 mg/cm$^{2}$ thick carbon target is shown as the shaded histogram
in Fig. \ref{fig:total}, together with the total spectrum of protons
from the plastic target measured by both telescopes. The thickness
of the carbon target was adjusted to match the thickness of the plastic
target in terms of energy loss. The {}``carbon'' background spectrum
was scaled by a factor of 1.98 to reflect the difference between the
integrated number of $^{8}$B ions accumulated during the main and
background runs, adjusted to account for the different number of carbon
atoms per cm$^{2}$ for the same energy loss in the carbon and plastic
targets. The polynomial fit to the {}``carbon'' background (shown
as a solid line in Fig. \ref{fig:total}) was then subtracted from
the spectrum of protons obtained with the plastic target.

\section{Structure of $^{9}$C in the Continuum Shell Model.}

The continuum of reaction states is an inseparable part of the $^{9}$C
structure. Thus, for the analysis and interpretation of the experimental
results we used the Continuum Shell Model (CSM) developed in Ref.
\cite{Volya2005}. The traditional shell-model (SM) Hamiltonian that
describes the internal evolution of the system is supplemented here
with the continuum of reaction states. For the model Hamiltonian we
use an s-p-sd-pf valence space with the WBP interaction \cite{Brown01}.
The Hamiltonian for protons in the continuum is given by a Woods-Saxon
potential. The parameters of this potential were taken from Ref. \cite{Schwierz06}.
The experimental value of the reduced mass of the $^{8}$B+$p$ system
was used. For the discussion here, the relevant set of channels includes
the $2^{+}$ ground state and $1^{+}$ first excited state of $^{8}$B.
These states are well reproduced by the above WBP interaction, but
the CSM reaction calculations were performed using the actual experimentally-known
$Q$-values. The results from the CSM calculation are presented in
Table \ref{tab:csm1}. This table also includes the spectroscopic
factors computed for the corresponding channels. 

The spectroscopic factors are defined here in a conventional way as
overlap integrals $\left\langle \psi_{SM}|\psi_{^{8}B}\otimes\psi_{p}\right\rangle $,
neglecting the effects of the continuum. Using these spectroscopic
factors, and the decay widths $\Gamma_{WS}$ resulting from the potential-model
calculation, a perturbative approach can also be used to estimate
the decay widths. The results of the perturbative calculation (not
shown) are extremely close to those from the full CSM calculation
given in the table. This agreement highlights the fact that CSM, by
construction, extends the traditional SM approach yet yields identical
results for bound states and almost unchanged results for isolated
narrow resonances as the conventional SM. To assess the role of the
continuum beyond the lowest order in perturbation, in the third column
we show the eigenstate reorientation angle $\Theta$, defined as $\cos\Theta=\langle\psi_{SM}|\psi_{CSM}\rangle$.
This angle reflects the relative change in the wave function due to
the presence of the continuum. For bound states, the angle is zero
confirming that SM and CSM eigenstates are identical. The second $3/2^{-}$
shows a noticeable deviation which can be explained by its interaction
through the continuum with a higher-lying $3/2^{-}$ state at 6.27
MeV in excitation having a width (determined by the CSM) of 3.6 MeV.
Despite their almost 2.6 MeV separation, the large widths allow these
two $3/2^{-}$ states to overlap leading to the noticeable changes
in the structure. The case of the $3/2^{+}$ state is similar.

It is well established that parametrization of a broad resonance by
the energy of its centroid and its width is particularly ambiguous
in the case of broad and/or overlapping resonances, as well as those
close to thresholds. The above case of the interference between $3/2^{-}$
states is a good example of this phenomenon. The interference effects,
while having a moderate effect on the parameters of the resonances,
nonetheless result in noticeable changes in the cross sections. In
Fig. \ref{fig:RCSM} (discussed below), the CSM was used to calculated
the inelastic $^{8}$B+p cross section, which at low energies is dominated
by the above mentioned pair of $3/2^{-}$ states. While the appearance
of the peaks is very consistent with a parametrization by energy and
width, the off-peak behavior reflects the significant role of interference.
The comparison with R-matrix calculations that include only a single
resonance is indicative of this phenomenon. (Further comparison of
cross section curves and theoretical methods is interesting but remains
beyond the scope of this paper.)

Given the large variety of available shell model interactions, we
estimate the theoretical uncertainty in our results by conducting
another calculation with a different interaction. The {}``PWT\char`\"{}
interaction \cite{Brown01} was used for the results given in Table
\ref{tab:csm2}. This older interaction includes only the p-shell.
The Woods-Saxon Hamiltonian describing the continuum remained unchanged
in this calculation.

\begin{table*}

\caption{\label{tab:csm1}Continuum Shell Model results for $^{9}$C performed
with the WBP interaction \cite{Brown01} in an s-p-sd-pf valence space.
Calculated excitation energies, widths, spectroscopic factors and
eigenstate reorientation angles are given for the first five states
in $^{9}$C. The known experimental value for the excitation energy
of the $1/2^{-}$ state \cite{Tilley2004} and the excitation energy
of the $5/2^{-}$ state measured in this work were used in the reaction
calculations.}

\begin{tabular}{ccccc|ccc|ccc}
\hline 
&
E$_{th}$&
E$_{exp}$&
$\Gamma$&
$\Theta$&
\multicolumn{3}{c|}{$\psi_{p}\otimes\psi_{^{8}B(gs)}$}&
\multicolumn{3}{c}{$\psi_{p}\otimes\psi_{^{8}B(1^{+})}$}\tabularnewline
\raisebox{1.5ex}{J$^{\pi}$}&
(MeV)&
(MeV)&
(MeV)&
deg&
S(p$_{3/2}$)&
S(p$_{1/2}$)&
$\Gamma_{WS}$&
S(p$_{3/2}$)&
S(p$_{1/2}$)&
$\Gamma_{WS}$ \tabularnewline
\hline 
3/2$^{-}$&
0.00&
0.00&
0.000&
0.0&
0.87&
0.00&
0.0&
0.18&
0.00&
0.0\tabularnewline
1/2$^{-}$&
1.4&
2.2&
0.027&
0.1&
0.18&
0.00&
0.15&
0.75&
0.00&
7.6E-3 \tabularnewline
5/2$^{-}$&
3.9&
3.6&
1.30&
5.1&
0.13&
0.59&
1.80&
0.00&
0.00&
0.59\tabularnewline
3/2$^{-}$&
4.1&
- &
1.32&
10.8&
0.08&
0.09&
3.19&
0.23&
0.47&
1.14\tabularnewline
\hline 
&
&
&
&
S(d$_{5/2}$)&
S(d$_{3/2}$)&
S(s$_{1/2}$)&
S(d$_{5/2}$)&
S(d$_{3/2}$)&
S(s$_{1/2}$)&
\tabularnewline
\hline 
3/2$^{+}$&
4.2&
- &
2.1&
16.0&
0.00&
0.01&
0.41&
0.05&
0.002&
0.03\tabularnewline
\hline
\end{tabular}
\end{table*}

\begin{table*}

\caption{\label{tab:csm2} Continuum Shell Model results for $^{9}$C performed
with the PWT interaction \cite{Brown01} in a p-shell-only valence
space.}

\begin{tabular}{ccc|ccc|ccc}
\hline 
&
E$_{x}$&
$\Gamma$&
\multicolumn{3}{c|}{$\psi_{p}\otimes\psi_{^{8}B(gs)}$}&
\multicolumn{3}{c}{$\psi_{p}\otimes\psi_{^{8}B(1^{+})}$}\tabularnewline
\raisebox{1.5ex}{J$^{\pi}$}&
(MeV)&
(MeV)&
S(p$_{3/2}$)&
S(p$_{1/2}$)&
S(tot)&
S(p$_{3/2}$)&
S(p$_{1/2}$)&
S(tot) \tabularnewline
\hline 
3/2$^{-}$&
0.00&
0.000&
0.95&
0.02&
0.97&
0.34&
0.03&
0.37 \tabularnewline
1/2$^{-}$&
2.2&
0.045&
0.30&
0.0&
0.30&
0.49&
0.01&
0.50\tabularnewline
5/2$^{-}$&
3.6&
1.44&
0.11&
0.68&
0.79&
0.03&
0.00&
0.03\tabularnewline
3/2$^{-}$&
4.1&
1.25&
0.01&
0.16&
0.17&
0.24&
0.38&
0.62 \tabularnewline
\hline
\end{tabular}
\end{table*}

\section{Inelastic background in the proton spectrum.}

\begin{figure}
\includegraphics[width=1\columnwidth]{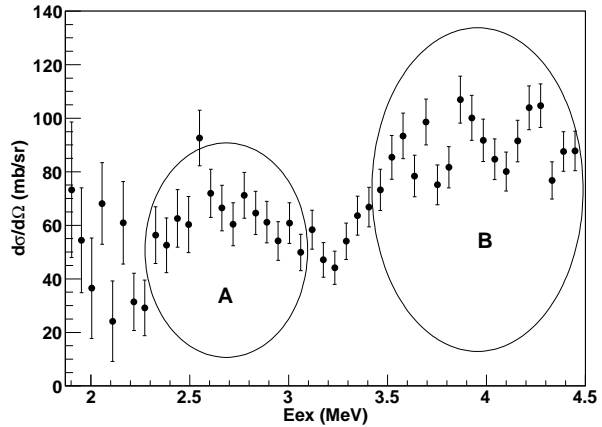}

\caption{\label{fig:cminel}Excitation function for resonance scattering of
$^{8}$B on protons at 164$^{\circ}\pm$7$^{\circ}$. The $^{9}$C
excitation energy is shown on the x-axis (this is the c.m. energy
plus the proton binding energy of 1.3 MeV).}
\end{figure}

The excitation function for the resonance scattering of $^{8}$B on
protons is shown in Fig. \ref{fig:cminel}. The background associated
with carbon in the target and with direct protons has been subtracted
as described in Sec.II above. Conversion into the center-of-momentum
(c.m.) system was performed individually for each bin of the histogram
by a computer code which takes into account the geometry of the experiment,
the integrated number of accumulated $^{8}$B ions, the energy losses
of the $^{8}$B and protons in the target, and the effective target
thickness for a specific bin. Two major features in the excitation
function, indicated by ellipses and labeled by letters A and B, are
apparent in Fig. \ref{fig:cminel}. These features could have been
associated with broad resonances at excitation energies $\sim$2.7
MeV and $\sim4$ MeV in $^{9}$C, but no states were reported or predicted
in $^{9}$C at these excitation energies \cite{Tilley2004}. The only
known excited state in $^{9}$C is a narrow (100 keV) 1/2$^{-}$ state
at 2.2 MeV which cannot be associated with either of these features.

More information is available on the level structure of the mirror
nucleus $^{9}$Li. The second excited state in $^{9}$Li, which has
a tentative spin-parity assignment of 5/2$^{-}$, has been observed
at an excitation energy of 4.3 MeV \cite{Tilley2004}. The analog
state is a good candidate for an explanation of the second peak in
the $^{8}$B+p excitation function (feature B in Fig. \ref{fig:cminel}).
However, the first peak (feature A) has no suitable counterpart in
the spectrum of $^{9}$Li. The large width of this peak ($\sim$500
keV) at 1.3 MeV above the threshold for proton decay indicates that
the wave function of this state should have a significant contribution
from the single particle configuration $\psi_{p}\otimes\psi_{^{8}B(gs)}$.
A state with such properties should have been observed in the recent
$^{8}$Li(d,p) experiment of Wuosmaa, \textit{et al.}, yet no states
between the known 1/2$^{-}$ state at 2.6 MeV and the state at 4.3
MeV were observed \cite{Wuosmaa2005}.

Various theoretical calculations also fail to offer any hint on what
the structure at 2.7 MeV might be. No states in $^{9}$Li/$^{9}$C
between the 1/2$^{-}$ and 5/2$^{-}$ states are predicted, either
by large basis no-core shell-model calculations \cite{Navratil1998},
or by Quantum Monte Carlo calculations \cite{Pieper2002}. (These
predictions have to be treated with caution since configurations that
allow positive parity states were not included). Continuum Shell Model
(CSM) calculations, described in the previous section, also predict
no states between the 1/2$^{-}$ and the 5/2$^{-}$ states. The considerations
discussed above do not completely rule out a state at 2.7 MeV in $^{9}$C,
but they make its existence very unlikely and stimulate the search
for another explanation of this feature.

The main disadvantage of the experimental technique used in this experiment
is the inability to distinguish between elastic and inelastic scattering.
Indeed, in the case of inelastic scattering, the recoil proton produced
in the process also can hit the detector and hence be mistaken for
an elastically-scattered proton. Normally, at low excitation energies,
the cross section for resonance elastic scattering is much higher
than that for inelastic scattering and the contribution from inelastic
processes can safely be ignored. In fact, this is almost the case
here. Estimations made using both CSM and R-matrix approaches (details
are given in the following section) produce a cross section for the
inelastic process which is one order of magnitude lower than that
for elastic scattering. However, since the excitation energy of the
first excited state of $^{8}$B is 0.77 MeV, which is 0.63 MeV above
the threshold of the $^{8}$B proton decay, an extra proton will be
produced in the inelastic scattering process. Moreover, due to the
inverse kinematics of the experiment, the extra proton will be focused
more toward forward angles, increasing the chance to hit the detectors.
The energy of this proton can be estimated using the following arguments.
The cross section for inelastic scattering is likely to increase toward
higher beam energies due to higher penetrability factors. Hence most
of the {}``inelastic'' protons are produced at the beginning of
the target. After inelastic scattering, the $^{8}$B recoils are focused
into a narrow cone ($\sim$6$^{\circ}$ with respect to the beam axis)
and have an energy of $\sim$3.0 MeV/A. The energy of protons produced
in the decay of the excited state of $^{8}$B is:

\begin{equation}
E_{p}=\frac{E_{^{8}B}}{8}+\frac{7}{8}Q+2\sqrt{\frac{7}{8}Q}\sqrt{\frac{E_{^{8}B}}{8}}\cos\theta,\label{eq:eplab}\end{equation}

where Q is the decay energy (0.63 MeV) and $\theta$ is the angle
between the momentum vectors of the proton and the $^{8}$B, which
has to be close to 0$^{\circ}$ or 180$^{\circ}$ in order for the
proton to hit the detector at 7$^{\circ}$. If $\theta$ is 0$^{\circ}$,
then E$_{p}$ is roughly 6 MeV. The energy loss of 6 MeV protons in
the 9 mg/cm$^{2}$ plastic target is $\sim$1 MeV. This leaves us
with 5 MeV protons in the detector. Note that only a limited range
of angles $\theta$ can produce a hit in the detector. Hence the protons
from the decay of the first excited state of $^{8}$B will peak at
around 5 MeV. This is exactly the energy of the first peak (A) in
the laboratory frame (see Fig. \ref{fig:total}). On the other hand,
if the angle $\theta$ is close to 180$^{\circ}$ the proton energy
will be $\sim$1.0 MeV and these protons will be stopped in the target.
Therefore, only one peak from the proton decay of the $^{8}$B first
excited state can be observed.

\begin{figure}
\includegraphics[width=1\columnwidth]{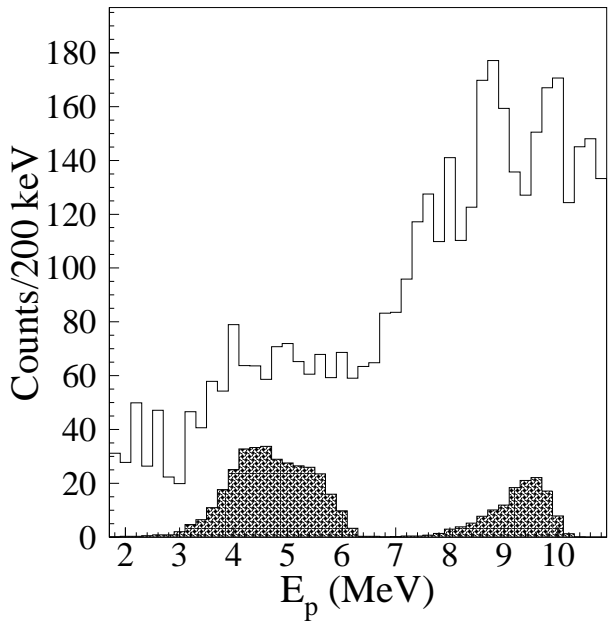}

\caption{\label{fig:inelastic}The proton spectrum in the laboratory system
with the carbon background subtracted. The Monte-Carlo simulation
of the inelastic process $^{1}$H($^{8}$B,p')$^{8}$B$^{*}$(1$^{+}$;0.77
MeV) is shown as a shaded histogram. The two peaks at 5 and 9 MeV
in the shaded histogram are associated with the proton decay of the
$^{8}$B excited state and the inelastically scattered proton, respectively.}
\end{figure}

A realistic Monte-Carlo simulation of the inelastic scattering $^{1}$H($^{8}$B,p')$^{8}$B$^{*}$(1$^{+}$;0.77
MeV) with subsequent proton decay of $^{8}$B was performed. The simulation
took into account the specific geometry of the experiment, the kinematics
of the process, energy losses of the $^{8}$B and protons in the target,
energy straggling, and multiple scattering (the GEANT 3.21 package
was used \cite{Brun94}). The result of this simulation is shown in
Figure \ref{fig:inelastic}. The first peak of the shaded histogram
in Fig. \ref{fig:inelastic} is associated with protons from the decay
of the excited $^{8}$B state, and the second peak corresponds to
the recoil protons inelastically scattered by $^{8}$B. It can be
seen from Fig. \ref{fig:inelastic} that {}``inelastic'' protons
can entirely account for the peak observed in the proton spectrum
at 5 MeV if the cross section for inelastic scattering of $^{8}$B
is $\sim$1/10 of the elastic cross section at 29 MeV. In the following
analysis, we will assume that this is the case and subtract the {}``inelastic''
proton contribution from the excitation function.

\section{R-matrix analysis}

\begin{figure}
\includegraphics[width=1\columnwidth]{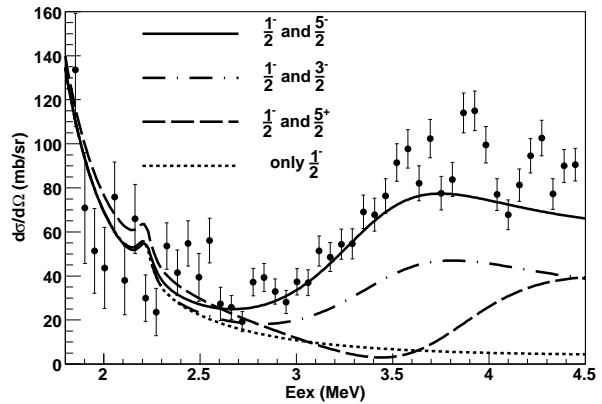}

\caption{\label{fig:fcomp}The excitation function of resonance elastic scattering
of $^{8}$B on protons after subtraction of {}``direct'', {}``carbon''
and {}``inelastic'' backgrounds. The solid curve is an R-matrix
fit with the known $1/2^{-}$ resonance plus a $5/2^{-}$ state at
an excitation energy of 3.6 MeV having a width of 1.5 MeV. The dotted
curve shows the contribution of the $1/2^{-}$ state alone. R-matrix
calculations for the $1/2^{-}$ state together with $3/2^{-}$ or
$5/2^{+}$ states are shown as dash-dotted and dashed curves, respectively. }
\end{figure}

The excitation function of resonance elastic scattering of $^{8}$B
on protons resulting from the subtraction of {}``direct'' proton
background, {}``carbon'' background and assumed {}``inelastic''
proton background, and converted into the c.m. system, is shown in
Figure \ref{fig:fcomp}. A two-channel multi-level R-matrix approach
was then applied for the analysis of the excitation function. Besides
the elastic channel, the first inelastic channel was included in the
R-matrix calculation. In order to make the R-matrix fit more realistic,
we start by deducing the reduced-width amplitudes from the CSM spectroscopic
amplitudes. Since these amplitudes have been calculated in the jj-coupling
scheme, they must be re-coupled into LS coupling as used in the R-matrix
approach. This can be accomplished using the 6$j$-symbols, according
to the following expression:

\begin{equation}
\begin{array}{lc}
\gamma_{\lambda c}= & \left(\frac{\hbar^{2}}{\mu_{c}a_{c}^{2}}\right)^{\frac{1}{2}}{\displaystyle \sum_{j}}A_{\lambda cj}(-1)^{^{I_{1}+I_{2}+\ell+J}}\\
 & \sqrt{2S+1}\sqrt{2j+1}\left\{ \begin{array}{ccc}
I_{1} & I_{2} & S\\
\ell & J & j\end{array}\right\} ,\end{array}\label{eq:rw}\end{equation}

where $c$ corresponds to a specific channel with a set of quantum
numbers J, $\ell$, S (the channel spin $S=I_{1}+I_{2}$), I$_{1}$
(the spin of the projectile), I$_{2}$ (the spin of the target), $\mu_{c}$
is the reduced mass, $a_{c}$ is the channel radius (4.5 fm), and
$A_{\lambda cj}$ is the CSM spectroscopic amplitude (the relative
phase was calculated in the CSM and defined by the sign of the $A_{\lambda cj}$).
The sum is taken over the relevant single-particle orbits ($j=1/2$
and $3/2$ for p-shell states). Reduced-width amplitudes, calculated
as described above, were then varied about the calculated values to
obtain the best fit to the data shown in Fig. \ref{fig:fcomp}.

It is clear that the known 1/2$^{-}$ first excited state of $^{9}$C,
at an excitation energy of 2.2 MeV and having a width of 100 keV \cite{Tilley2004},
cannot account for the large cross section observed at higher energy
(dotted curve in Fig \ref{fig:fcomp}). Introduction of a broad 5/2$^{-}$
state at an excitation energy of $\sim$3.6 MeV produces a reasonable
agreement between the R-matrix calculation and the experimental data
(solid curve in Fig. \ref{fig:fcomp}). The excitation energy and
width of the assumed $5/2^{-}$ state are $3.6\pm0.2$ MeV and $1.4\pm0.5$
MeV, respectively. No other spin-parity assignment for the state at
3.6 MeV agrees with the experimental data. The cross section for the
3/2$^{-}$ state is too small (dash-dotted curve in Fig. \ref{fig:fcomp}),
and the positive parity states, with $\ell$=0 dominant partial wave,
produce a dip in the excitation function due to destructive interference
(dashed curve in Fig. \ref{fig:fcomp}).

\begin{figure}
\includegraphics[width=1\columnwidth]{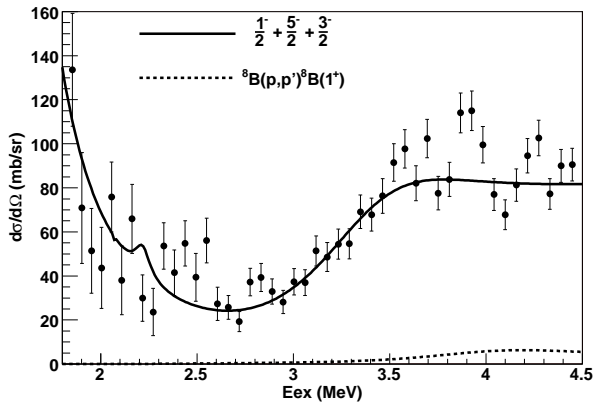}

\caption{\label{fig:finel}Solid curve is an R-matrix fit of the $^{8}$B+p
excitation function which includes $1/2^{-}$, $5/2^{-}$ and $3/2^{-}$
resonances. Dotted curve is the calculated excitation function of
the resonance inelastic scattering (it is mostly due to the $3/2^{-}$
resonance).}
\end{figure}

Two problems can be identified in the description of the excitation
function as a combination of only two states, $1/2^{-}$ and $5/2^{-}$.
First, the measured cross section is still somewhat higher at $\sim$4
MeV than can be accounted for by the $5/2^{-}$ state in $^{9}$C.
Second, the $5/2^{-}$ state is essentially a single-particle state
and decays predominantly to the ground state of $^{8}$B. If no other
states are introduced, the cross section for inelastic scattering
{[}$^{1}$H($^{8}$B,p')$^{8}$B(1$^{+}$)] would be too small to
account for the {}``inelastic'' peak at 5 MeV in the laboratory
proton spectrum (Fig. \ref{fig:inelastic}). Both problems can be
resolved at once if an additional state with substantial contribution
from the $^{8}$B(1$^{+}$)+p configuration is introduced. The CSM
calculations predict two states in close proximity to the $5/2^{-}$
state (see Table \ref{tab:csm1}). These are $3/2^{-}$ and $3/2^{+}$
states at 4.1 and 4.2 MeV. The wave function of the $3/2^{-}$ state
has a large inelastic component (Table \ref{tab:csm1}). Hence, the
introduction of this state can potentially fix both problems by increasing
the cross section at $\sim4$ MeV and explaining the observed {}``inelastic''
peak. The solid curve in Fig. \ref{fig:finel} shows an R-matrix fit
with $1/2^{-}$, $5/2^{-}$ and $3/2^{-}$ states. The dashed curve
in Fig. \ref{fig:finel} represents the inelastic excitation function
due to the $3/2^{-}$ state. The best-fit R-matrix parameters for
all three states are shown in Table \ref{tab:rmatrix} (the values
given in parenthesis are reduced-width amplitudes calculated from
CSM spectroscopic amplitudes using the WBP \cite{Brown01} interaction,
according to the expression \ref{eq:rw}). The parameters for the
$1/2^{-}$ state (energy eigenvalue and elastic reduced-width amplitude)
were adjusted to reproduce the experimental excitation energy and
width \cite{Tilley2004}. The parameters for the $3/2^{-}$ state
were derived from CSM and were not varied. As follows from the Table
\ref{tab:rmatrix}, no significant modifications of the calculated
CSM reduced-width amplitudes (except for the elastic reduced-width
amplitude of the $1/2^{-}$ state) were necessary to fit the experimental
data.

\begin{table}

\caption{\label{tab:rmatrix}Excitation energies, energy eigenvalues, widths
and reduced-width amplitudes of the resonances deduced in $^{9}$C.}

\begin{tabular}{lcccccc}
\hline 
&
$\ell$&
S&
B$_{c}$&
&
&
\tabularnewline
J$^{\pi}$&
&
&
&
$\frac{1}{2}^{-}$&
$\frac{5}{2}^{-}$&
$\frac{3}{2}^{-}$ \tabularnewline
\hline 
E$_{ex}$ (MeV)&
&
&
&
2.22&
3.6&
4.1 \tabularnewline
E$_{\lambda}$ (MeV)&
&
&
&
0.61&
3.45&
4.15 \tabularnewline
$\Gamma$ (MeV)&
&
&
&
0.10&
1.4&
1.3 \tabularnewline
\hline 
&
1&
$\frac{3}{2}$&
-1.2&
1.15 (0.65)&
0.33 (0.47)&
0.17 (0.17)\tabularnewline
\raisebox{1.5ex}{p+$^{8}$B(g.s.)}&
1&
$\frac{5}{2}$&
-1.2&
-&
-1.34 (-1.20)&
0.59 (0.59)\tabularnewline
&
1&
$\frac{1}{2}$&
-1.2&
1.25 (1.25)&
-&
-0.15 (-0.15)\tabularnewline
\raisebox{1.5ex}{p+$^{8}$B(1$^{+}$)}&
1&
$\frac{3}{2}$&
-1.2&
0.42 (0.42)&
0.00 (0.00)&
1.27 (1.27)\tabularnewline
\hline
\end{tabular}
\end{table}

\section{Discussion}

The main result of this work is a firm identification of the $5/2^{-}$
resonance in $^{9}$C. Its excitation energy is 3.6$\pm$0.2 MeV and
its width is 1.4$\pm$0.5 MeV. This state is a feature of all theoretical
calculations, \textit{ab initio} \cite{Pieper2002,Navratil1998} and
shell-model alike, which seem to agree on the structure and excitation
energy of the state. These predict that it should be in the vicinity
of 3.5 MeV and have a single particle structure with spectroscopic
factor of about 0.8. The experimental width of the $5/2^{-}$ state
is in very good agreement with the CSM width (1.30 MeV). Based on
this comparison, the experimental single-particle spectroscopic factor
of the $5/2^{-}$ state is 0.77$\pm$0.25. This result is also in
very good agreement with the recent experimental result of Wuosmaa,
\textit{et al.} \cite{Wuosmaa2005}, in which the spectroscopic factor
of the mirror state in $^{9}$Li was measured to be 0.93$\pm$0.2
using the $^{8}$Li(d,p) reaction.

Unfortunately, the statistics in the present experiment were not sufficient
to observe the weak $1/2^{-}$ first excited state. The direct observation
of this state in the $^{8}$B+p excitation function would allow for
an accurate measurement of its width. At the moment, it seems that
theoretical calculations (especially shell model) tend to underestimate
the single particle spectroscopic factor of this state. The width
of this state in $^{9}$C was determined in a single, low-statistics
experiment using the $^{12}$C($^{3}$He,$^{6}$He) reaction \cite{BENENSON1974},
and the measured value of 100$\pm$20 keV indicates that the spectroscopic
factor of the state is $\sim$0.67$\pm$0.15 based on a comparison
to the CSM width given in Table \ref{tab:csm1}. A similar result
was obtained for the spectroscopic factor of the $1/2^{-}$ state
in $^{9}$Li using the $^{8}$Li(d,p) reaction \cite{Wuosmaa2005}.
However, the theoretical predictions for the spectroscopic factor
(which range from 0.17 to 0.5 depending on the model and interaction
used in the SM) are consistently lower than the experimental value.
Hence, it would be of interest to verify the result of the previous
experiment and improve the experimental accuracy for the value of
the width of this state.

\begin{figure}
\includegraphics[width=1\columnwidth]{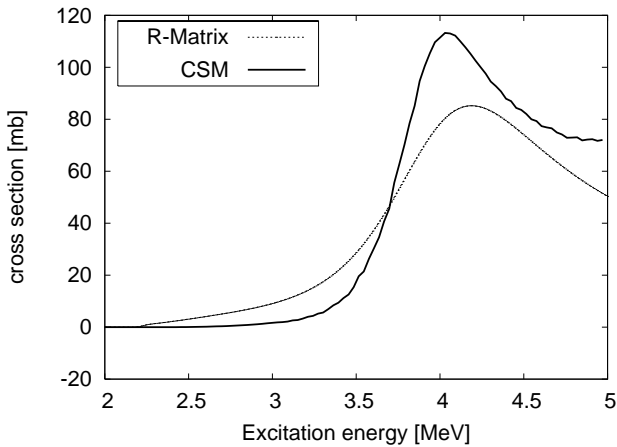}

\caption{\label{fig:RCSM}The total cross section of the inelastic scattering
process $^{8}$B(p,p')$^{8}$B(1$^{+}$). The dotted curve was calculated
using the R-matrix approach with resonance parameters given in Table
\ref{tab:rmatrix}. The solid curve is the CSM calculation, which
automatically take into account the influence of higher- lying resonances.}
\end{figure}

In addition to the known $1/2^{-}$ state and the proposed new $5/2^{-}$
state, the experimental data indicate the presence of an additional
state or states at or above 4 MeV with strong decay branch(es) into
the first excited state of $^{8}$B. The $3/2^{-}$ state predicted
by CSM at 4.1 MeV (Table \ref{tab:csm1}) seems to be a good candidate.
The inclusion of the $3/2^{-}$ state improved the fit and provided
for an explanation of the {}``inelastic'' peak at 5 MeV of proton
laboratory energy. The inelastic cross section due to the $3/2^{-}$
state (shown in Figure \ref{fig:finel}) was calculated using an R-matrix
approach with parameters for the $3/2^{-}$ state derived from the
CSM as described in the previous section. This cross section can also
be calculated directly in the CSM approach in which all the resonances
(including resonances at higher excitation energy) are included automatically
with correct interference. The comparison of the total $^{8}$B(p,p')$^{8}$B(1$^{+}$)
inelastic cross section, calculated using the R-matrix approach and
the CSM, is shown in Fig. \ref{fig:RCSM}. The magnitude and the shape
of the total inelastic cross section is similar in both calculations.
This comparison is instructive in many ways. First, it shows that
the R-matrix calculation with reduced-width amplitudes derived from
the CSM spectroscopic amplitudes using Eq. \ref{eq:rw} produces a
cross section in the vicinity of the $3/2^{-}$ resonance which is
similar to the one calculated directly in the CSM . Second, it indicates
that, while the $3/2^{-}$ resonance included in the R-matrix calculation
determines the inelastic cross section at $\sim$4 MeV, the shape
of the inelastic cross section is influenced by higher- lying resonances
(especially by the next $3/2^{-}$ state). The CSM automatically takes
this into account, providing a realistic estimate of the influence
of background resonances and accounting for the possible interference
of this background with the main resonances. This feature of the CSM
is important for future studies of the properties of exotic nuclei
in resonance scattering. Unfortunately, due to the limited excitation
energy range, relatively low statistics, and inability to separate
elastic from inelastic processes, the inelastic cross section cannot
be accurately extracted from the present experimental data and the
actual parameters of the $3/2^{-}$ state cannot be determined.

In Ref. \cite{GOLOVKOV1991}, the authors claim that they observe
a resonance in $^{9}$C at 3.3 MeV in the $^{12}$C($^{3}$He,$^{6}$He)
reaction. Based on the assumption of mutual correspondence between
the 4.3 MeV state in $^{9}$Li and the observed state in $^{9}$C,
they concluded that the state cannot be $5/2^{-}$, arguing that the
observed Thomas-Ehrman shift is too large for an $\ell$=1 state.
A $5/2^{+}$ spin-parity assignment was proposed instead. However,
close examination of the $^{6}$He spectrum in Fig. 1 of Ref. \cite{GOLOVKOV1991}
reveals that, due to low statistics and poor experimental resolution
($\sim500$ keV), the spectrum allows for a different interpretation
and also does not contradict assignment of a broad state at 3.6 MeV.
Moreover, the value for the Thomas-Ehrman shift of the $5/2^{-}$
state in $^{9}$Li/$^{9}$C can be estimated from a simple potential
model assuming a pure single-particle nature for this state. For example,
a Woods-Saxon potential with common parameters ($r_{\circ}=1.21$
fm, $a=0.65$ fm) and depth adjusted to reproduce the c.m. energy
of the 4.3 MeV state in $^{9}$Li gives 0.5 MeV for the Thomas-Ehrman
shift of the $\ell$=1 state. This value is a factor of four bigger
than the one assumed in Ref. \cite{GOLOVKOV1991} and suggests that
the excitation energy of the $5/2^{-}$ state in $^{9}$C should be
$\sim$3.8 MeV. This value is in very good agreement with the results
of this work, lending further support to a $5/2^{-}$ spin-parity
assignment for the resonance in question.

\section{Conclusion}

The excitation function for resonance scattering $^{8}$B+p was measured
using the thick-target inverse-kinematics technique. A $^{9}$C excitation
energy range from 1.8 MeV to 4.5 MeV was covered. One new state in
$^{9}$C was identified at an excitation energy of 3.6$\pm$0.2 MeV,
having a width of 1.4$\pm$0.5 MeV. R-matrix analysis of the excitation
function allows for a unique $5/2^{-}$ spin-parity assignment to
this state. It has a single-particle nature with a spectroscopic factor
of 0.77$\pm$0.25, consistent with theoretical predictions of the
\textit{ab initio} models and CSM calculations, and also with recent
experimental results for the presumed mirror state in $^{9}$Li \cite{Wuosmaa2005}.
The measured excitation function indicates the existence of higher-lying
states with strong inelastic decay branches. A new measurement with
higher statistics, broader excitation energy and angular range coverage
is highly desirable. The new experiment must be designed in a way
that allows for separation of elastic and inelastic scattering.

Using the CSM as a part of the analysis, we attempted to take a step
beyond the typical perturbative approach based on spectroscopic factors
and R-matrix analysis. In the case of the newly-discovered resonances,
due to their isolated nature, the role of the continuum appears to
be reasonably well described by perturbation theory. However, the
onset of physics that demands the use of a unified structure-reaction
approach is clearly indicated. As an example, the experimental analysis
hinges on tracking the contribution from an inelastic channel that
appears to be dominated by overlapping $3/2^{-}$ resonances. The
CSM was used to compute the cross section for this process, and comparison
with the R-matrix fit indicates a non-trivial nature for new physics
on the structure/reaction border that is successfully captured by
the novel CSM technique.

\begin{acknowledgments}
The authors are grateful to Prof. Goldberg for helpful discussions.
This work was supported by the National Science Foundation under Grant
Nos. PHY04-56463 and PHY03-54828, and by U.S. Department of Energy
Contract No. DE-FG02-92ER40750.

\bibliographystyle{plain} \bibliographystyle{plain} \bibliographystyle{plain}
\bibliography{9c}

\end{acknowledgments}

\end{document}